\documentclass[twocolumn]{emulateapj}

\usepackage{epsfig}
\usepackage{amsmath}
\usepackage{natbib}

\begin{document}

\title{The Einstein ring 0047-2808 Revisited: A Bayesian Inversion}
\author{B. J. Brewer and G. F. Lewis}
\email{brewer@physics.usyd.edu.au}
\email{gfl@physics.usyd.edu.au}
\affil{Institute of Astronomy, School of Physics, A28, 
University of Sydney, NSW 2006, Australia}

\begin{abstract}
In a previous  paper, we outlined a new  Bayesian method for inferring
the  properties of extended  gravitational lenses,  given data  in the
form  of resolved  images.  This  method  holds the  most promise  for
optimally  extracting  information  from  the observed  image,  whilst
providing reliable uncertainties in all parameters. Here, we apply the
method  to the  well  studied optical  Einstein  ring 0047-2808.   Our
results are in broad agreement with previous studies, showing that the
density profile of the lensing  galaxy is aligned within a few degrees
of  the light  profile,  and  suggesting that  the  source galaxy  (at
redshift 3.6) is a  binary system, although  its size is  only of
order 1-2 kpc. We also find  that the mass of  the elliptical lensing
galaxy  enclosed  by  the  image  is  (2.91$\pm$0.01)$\times$10$^{11}$
M$_{\sun}$. Our method is able to achieve improved resolution for the source reconstructions, although we also find that some of the uncertainties  are greater than has been found in previous analyses, due to the inclusion of extra pixels and a more general lens model.
\end{abstract}

\keywords{gravitational lensing -- methods: data analysis -- methods: statistical}

\maketitle

\newcommand{\BL}{BL06}

\section{Introduction}
Gravitational  lenses have  long  held the  promise  of being  natural
telescopes, probing  structural scales smaller than  the resolution of
current technologies  while simultaneously revealing  the distribution
of   matter  within   the   lensing  galaxy   \citep[For  reviews   of
gravitational   lensing,  see][]{1992grle.book.....S,kochreview}.   In
practice, this promise can only  be truly realised in systems in which
the lensed images are extended,  as these can offer substantially more
constraints than multiply imaged  point-like quasars. For this reason,
significant  attention has  been paid  to extended  sources  lensed by
intervening  galaxies  to produce  roughly  circular  images known  as
Einstein rings \citep[][]{2001ApJ...547...50K}.

A long-standing  issue has  focused upon the  question of what  is the
{\it  best} approach  to invert  such extended  gravitationally lensed
images, providing the unlensed source brightness distribution and mass
in the lensing  galaxy. Addressing this question has  spawned a number
of   seemingly   distinct   methods,    such   as   the   ring   Cycle
\citep{1989MNRAS.238...43K},           Semi-Linear           Inversion
\citep{2003ApJ...590..673W}        and        Genetic       Algorithms
\citep{2005PASA...22..128B}. Given the  extended nature of the source,
reconstructions have been based upon  a pixellated source plane, so as
to make minimal assumptions about the source brightness distributions.
However, due  to the  ill-posed nature of  such an  inversion, authors
have tended to use low  numbers of pixels, and/or regularization. In a
previous contribution,  we demonstrated  that other approaches  can be
unified in terms of a Bayesian interpretation, with each corresponding
to  the use of  particular, often  unjustified, assumptions  about the
nature of the  source \citep[][hereafter \BL]{ours}. Furthermore, \BL\
presented a general Bayesian approach to the question of gravitational
lens inversion,  using realistic prior distributions  and Markov Chain
Monte Carlo (MCMC) methods to recover the properties of the lensing system.

In this present contribution, the approach detailed in \BL\ is applied
to   the   well-studied   optical   Einstein   ring   0047-2808.    In
Section~\ref{image},  the  details  of   this  system,  and  the  data
employed,  are  presented,  while Section~\ref{approach}  details  the
approach to  the problem.  Section~\ref{results}  outlines the results
of  this study,  with  a  comparison to  other  techniques, while  the
conclusions are presented in Section~\ref{conclusions}.

\section{ER 0047-2808}\label{image}

\subsection{Background}
The optical Einstein ring  0047-2808 was identified serendipitously in
a  survey  of  massive  elliptical  galaxies  at  $z\sim0.5$  via  the
identification of an anomalous  emission line, with subsequent imaging
revealing a ring of high redshift ($z=3.6$) emission superimposed upon
a   $z=0.49$  galaxy  \citep{1996MNRAS.278..139W,1999A&A...343L..35W}.
Initial estimates suggested that the  source was magnified by a factor
of $\sim17$ and hence detailed spectroscopy is able to probe this high
redshift,       kiloparsec       scale       star-forming       galaxy
\citep{1998MNRAS.299.1215W}.

\subsection{Observations}
Since its discovery  \citep{1996MNRAS.278..139W}, the optical Einstein
ring 0047-2808 has been the subject of several different studies whose
goal was  a full  lensing inversion and  reconstruction of  the source
brightness           distribution          \citep{1999A&A...343L..35W,
2003ApJ...583..606K,2005MNRAS.360.1333W,2005ApJ...623...31D}.       The
more recent of these have  focused upon observations obtained with the
Wide  Field  Planetary  Camera  2  (WFPC2) onboard  the  Hubble  Space
Telescope and given its superior resolution over previous ground-based
observations, this data is the subject of this current contribution.

\begin{figure}
\begin{center}\includegraphics[width = 0.45\textwidth]{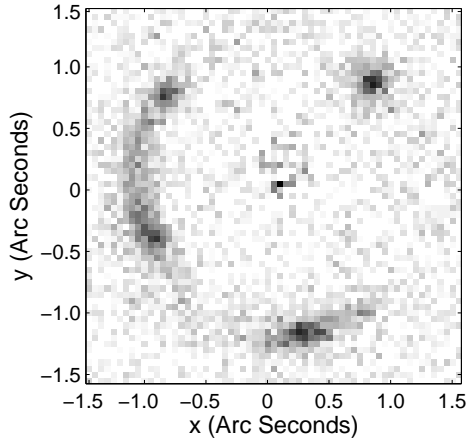}
\end{center}
\caption{The HST/WFPC2  image of  the optical Einstein  ring 0047-2808
employed in this study. The image has a side length of 3.05 arcseconds
and  comprises of  61x61  pixels.  As  noted  in the  text, the  light
distribution   of   the   foreground   galaxy  has   been   subtracted
\citep[see][]{2005MNRAS.360.1333W}; this foreground light distribution
increases the noise in the central regions of this image.\label{data}
}
\end{figure}

The raw pixel scale of the  WFPC2 CCD used in imaging 0047-2808 is 0.1
arcseconds,  but the image  used in  this study  is comprised  of four
interlaced  dithered  images,  giving   a  pixel  resolution  of  0.05
arcseconds    \citep[see][for    details]{2005MNRAS.360.1333W}.    The
foreground  lensing  galaxy has  been  subtracted,  after fitting  its
brightness  distribution with  a Sersic  profile.  A  region  of 61$\times$61
pixels,  centred on  the  resulting lensed  image  and encompassing  a
region  of 3.05$\times$3.05 arcseconds,  was extracted  and employed  in this
study; this image is presented in Figure~\ref{data}.

As with  \citet{2005MNRAS.360.1333W}, a noise frame  covering the same
region and including  the various noise components (such  as that from
the light distribution  of the lensing galaxy) was  also employed.  It
should be noted  that this study did not  recreate the image dithering
procedure that was  undertaken with the observed system,  rather it is
assumed  that  the image  presented  in  Figure~\ref{data}  is a  true
representation  of  this  lensing  system  at  a  resolution  of  0.05
arcseconds. To this end, the  point spread function was generated with
the TinyTim algorithm to match this resolution scale \citep{krist}. Although in reality the noise values between neighbouring pixels are correlated, we will use an uncorrelated likelihood function, which is computationally much more managable and will lead to slightly more conservative results.

\section{approach}\label{approach}

\subsection{Parameter Estimation}
The basis  of our  method was  presented in \BL.   For a  more general
introduction to  Bayesian methods, including Markov  Chain Monte Carlo
techniques, see the  textbook by \citet{2005blda.book.....G}. Assuming
a particular  form for the  lens model, and background  information or
assumptions $I$, we write  down the joint probability distribution for
the lens  and source parameters  (denoted collectively by $L$  and $s$
respectively)  given  the  data  $D$  and the  prior  information  and
assumptions $I$:

\begin{equation}
p(SL|DI) = \frac{p(SL|I)p(D|SLI)}{p(D|I)}
\end{equation}

The denominator does not depend  on the source and lens parameters, so
is  part  of the  normalisation  constant  for  the joint  probability
distribution. The  first factor  in the numerator  is the  joint prior
probability density of the lens  and source parameters, and the second
factor  is the likelihood  function (probability  density of  the data
that  was  actually observed,  given  the  parameters). Commonly,  the
source  is pixellated into  $m$ pixels  and is  described by  a vector
$\mathbf{s}$  of pixel  intensities, and  the observed  image  is also
pixellated  (with  $n$  pixels),   and  is  represented  by  a  vector
$\mathbf{O}$.   In  this case  the  image  predicted  by a  source  is
calculated by  a matrix multiplication: $\mathbf{I =  Ls}$. The matrix
$\mathbf{L}$ depends on  the lens parameters $L$ and  the point spread
function.  Since  it is  always possible that  the noise  level values
\{$\sigma_i$\}  are not  completely  reliable, we  also introduced  an
extra  noise  parameter  $\sigma$,  to  be  estimated  from  the  data
\citep{2005ApJ...631.1198G}. If a good fit is possible with our model,
$\sigma$ will be estimated to be small. If a good fit is not possible,
$\sigma$ will be  estimated to be large, and  the uncertainties in all
other  inferred parameters will  also be  increased. Thus, we can use the 
inferred value of $\sigma$ as a kind of alternative-free test of whether 
our lens model is adequate. Under  the usual
assumptions (a Gaussian probability distribution for the error in each
pixel of the image, and  logical independence of the prior information
about the lens and the source), we have the PDF

\begin{eqnarray}
p(\mathbf{s}L\sigma|DI) & \propto &  
p(\mathbf{s}|I)p(L|I)p(\sigma|I)p(\mathbf{O}|{s}L \sigma I) \nonumber \\
& \propto & p(\mathbf{s}|I)p(L|I)p(\sigma|I) \times  \nonumber 
\end{eqnarray}
\vspace{-0.5cm}
\begin{eqnarray}
\left[\prod_{i=1}^n\frac{1}{\sqrt{\sigma^2 + \sigma_i^2}}\right]
e^{-\frac{1}{2}\sum_{i=1}^n\frac{(O_i - \sum_{j=1}^mL_{ij}s_j)^2}{\sigma^2 + \sigma_i^2}}
\end{eqnarray}

where the right  hand side depends on the  lens parameters through the
matrix $\mathbf{L}$ in  some complicated nonlinear fashion. Therefore,
there is no hope of finding marginal distributions analytically, so we will use a  Markov Chain Monte  Carlo algorithm to generate  random samples
from this posterior distribution.

\subsection{Lens Model}
For  its  convenient  properties,   we  chose  an  softened  power-law
elliptical     potential     (SPEP)     for     the     lens     model
\citep{1998ApJ...502..531B}. Elliptical  potentials have the advantage
of  faster computation of  the deflection  angles than  for elliptical
projected mass  distributions, and great  flexibility in the  range of
possible lenses is  achievable with a handful of  parameters. 
Hopefully, this flexibility is sufficient to ensure that we won't make any 
overconfident conclusions caused by the fact that real lenses are not 
exactly described by some parametric model. The primary reason for choosing this model was computational speed, which is an 
important consideration because our MCMC algorithm requires a lot of 
likelihood evaluations.

The 2-D lensing potential we used was

\begin{equation}
\phi(x,y) = \frac{b}{\gamma}R^\gamma
\end{equation}
where
\begin{equation}
R = \sqrt{r_c^2 + x^2q + y^2/q}
\end{equation}

The gradient of $\phi$ gives the deflection angles as
\begin{equation}
\alpha_x(x,y) = \frac{\partial\phi}{\partial x} = (bqx)R^{\gamma - 2}
\end{equation}
\begin{equation}
\alpha_y(x,y) = \frac{\partial\phi}{\partial y} = (by/q)R^{\gamma - 2}
\end{equation}

The parameter $b$ is the overall  strength of the lens, and $q$ is the
ratio of the  ellipse's major and minor axes. $q  \sim 1$ indicates an
approximately  circularly  symmetric  lens. Furthermore,  $(x,y)$  is
position in  the lens plane, in  a coordinate system  aligned with the
lens and centred at the centre  of the lens. Thus, in fitting the lens
to  an observed image,  we have  7 free  parameters: $b$,  $q$, $(x_c,
y_c)$  (the position of  the centre  of the  lens), $\gamma$,  a slope
parameter,  the   core  radius  $r_c$  and  $\theta$,   the  angle  of
orientation of  the lens  mass distribution with  respect to  the axes
defined  by  the image.  If $q=1$,$\gamma=1$ and $r_c=0$, this model reduces to the common isothermal sphere model, for which $b$ is the Einstein Radius. It  is  not  clear whether  $(x_c,y_c)$  were
considered  known a priori  in previous  investigations; here  we will
consider them partially known  by assigning a weakly informative prior
probability density. Using the scaled lens equation with source and image plane positions measured in arc seconds corresponds to a choice of the units of mass, as being that which would make the Einstein radius equal to 1 arc second.

It is usually the  case that an image can be fitted  with a variety of
differently parametrized  lens models.  However, for most  purposes we
are  more  interested  in  the  values of  the  parameters  themselves
(e.g. how much  mass is there, how elliptical  is the density profile,
is it aligned  with the light profile of the  lensing galaxy, etc) and
little  would  be  gained  by  comparing  different  choices  for  the
parameterisation of the mass model.  However, we suspect that, if very
high  resolution images were  available, simple  lens models  would be
inadequate  and we  would  need  to use  a  nonparametric mass  model,
similar to that used by \citet{marshall} for weak lensing. It is difficult to determine at what point this would be neceessary, however, if our ``extra noise'' parameter $\sigma$ was inferred to be large then this would be a definite indication that a different mass model is required.

\subsection{Lens Parameter Priors}
In  order for  the problem  to have  a definite  solution, we  need to
introduce prior  probability distributions  for the parameters  of the
lens model, to  describe our prior knowledge or  ignorance about their
values. In  principle, we could  consider the image of  the foreground
galaxy as  providing prior information,  however we did not  carry out
any  sophisticated  analysis  along  these lines,  because  the  final
results   depend  only  very   weakly  on   the  specific   choice  of
functions. The prior probability  densities we used for each parameter
are  shown  in  Table~\ref{lenspriors}.  These are  only  intended  to
capture vague  prior information  about the range  of values  that the
parameter could plausibly take. For  the position of the centre of the
lens model, we  assumed this was near the centre  of the light profile
of the  foreground galaxy, but  the large prior standard  deviation of
0.2 means that this is a weak assumption.

\begin{table}
\begin{center}
\caption{Prior  probability densities for  the lens  model parameters,
and also the  extra noise parameter $\sigma$.}\label{lenspriors}
\begin{tabular}{lcc}
\hline Parameter & Prior Distribution  \\
\hline $b$ & Exponential, mean 1''$^{1-\gamma}$\\
$q$ & Normal, mean 1'', SD 0.2'' \\
$r_c$ & Exponential, mean 0.5'' \\
$\gamma$ & Normal, mean 1, SD 0.2 \\
$x_c$ & Normal, mean 0.092'', SD 0.2'' \\
$y_c$ & Normal, mean 0.164'', SD 0.2'' \\
$\theta$ & Uniform, between 0 and $2\pi$ \\
$\sigma$ & Jeffreys' uninformative prior $\propto 1/\sigma$
\end{tabular}
\medskip\\
\end{center}
\end{table}

\subsection{Source Prior}
Since  the source  may have  complex structure,  we need  to introduce
enough pixels  to capture  this in  detail.  We chose  to use  a 48x48
pixel grid  to represent  the source.  The  range of the  source plane
that was  covered was 0.6 arc  seconds across, so  the resolution over
the source plane was a factor  of 4 greater than the resolution of the
image.  If the image strongly constrains part of the source, this will
be  reflected in  the posterior  samples. This  may or  may  not occur
(depending on  the quality  of the data),  but using large  numbers of
pixels at  least keeps  the option  open.  Since we  are not  using an
optimisation based approach, overfitting does not occur.

In \BL,  we discussed the shortcomings of  regularization based priors
for  astronomy,  and suggested  simple  priors  which  should be  more
realistic. In this paper, we use a prior which is somewhat more complex than that described in \BL. The  prior probability  distribution describing the  range of
source  models   which  we  consider  plausible   was  constructed  as
follows. Since the source is mostly blank, rather than trying to infer
each  source pixel  value  directly,  we imagine  that  the source  is
generated by ``monkeys'' throwing  $N$ ``atoms'' of intensity onto the
(initially blank)  source plane.  Each  atom has four attributes: An
intensity $B$, a discrete position $(i,j)$, indicating which pixel
the atom landed in, and a width parameter. The prior probability density of the intensity $B$
of an  atom is  taken as  exponential with mean  $\mu$ (chosen to match the typical brightness scale of the image), and  the prior
distribution of the discrete position $(i,j)$ of each atom is uniform.
This   is   similar to  the   Massive   Inference   prior  suggested   by
\citet{1998mebm.conf....1S} for  positive, additive distributions such
as  surface brightness  distributions. It  has the  desirable property
that the  prior probability distribution for the  amount of integrated
light over any macroscopic region R of the source plane is independent
of how we choose to subdivide that region into pixels.

Allowing  the atoms  to have  variable size  allows us  to reconstruct
plausible  astrophysical sources with  only a  small number  of atoms,
negating the problems caused by an entropic prior, which tends to make
the  reconstructed  sky  too   bright,  possibly  also  causing  other
parameters to be  inferred incorrectly (\BL). MEM may  be suitable for
photographs, but for astronomy, MassInf is more appropriate. Also, there 
is some prior expectation that the source should contain correlations, so 
that bright pixels are more likely to be near other bright pixels. To take 
this into account, we allowed three different types of atoms to make up 
the source - single pixel sized atoms, and two ``fuzzier'' types of atoms 
that spread light over several pixels. The effect of this variable atom size is to reduce the impact of our particular choice of pixellation scale, by allowing finer structure in those parts of the source where it is justified by the data. We chose to discretize into three distinct types of atom to reduce computational load, because if we had allowed the atom width to be a continuous variable it would have taken much more computation to produce the pixellated source for lensing.

In our  MCMC simulations, we  modified the
source by manipulating the  atoms (positions, intensities, widths, and 
number) directly, rather  than by modifying  source pixel values directly.  This helps,
because an  increased acceptance rate  can be achieved  using proposal
transitions that move  an atom from one pixel  to its neighbour; since
this  will only  have  a slight  effect  on the  predicted image,  the
acceptance probability will be  fairly high. Also, most proposal moves
for the Metropolis algorithm will be concentrated around modifying the
bright parts  of the source (where  the atoms are), so  little time is
wasted adjusting pixels where the source is mostly blank.

\subsection{MCMC Sampling}
A feature  of the lens inversion problem  is the fact that  it is much
faster  (by a factor  of $\sim$  50) to  compute the  image for  a new
source, with a fixed lens model, than it is to compute the image for a
new lens parameter set. This  is because, to calculate the image after
modifying  only  the source,  a  matrix  multiplication  (by a  sparse
matrix)  is required.  However, to  calculate  the new  image after  a
change in  the lens  parameter values requires  that we  calculate the
sparse lensing/blurring matrix $\mathbf{L}$, which requires a large ray tracing
calculation, firing  100 rays  per pixel in  order to get  an accurate
approximation to the lensing  matrix. A modified Metropolis method has
been  developed  by \citet{neal}  which  is  designed  to improve  the
sampling when some  variables are ``slow'' and others  ``fast'', so we
implemented     this    in     our     code.    Parallel     tempering
\citep{2005blda.book.....G}   was  also   used  to   ensure  efficient
exploration of the parameter space. The MCMC simulation was run several times with different random number seeds, in order to check convergence. The results were deemed to be reliable enough if the variance of the error bars returned was significantly less than their size, which was the case.

\section{Results}\label{results}
\subsection{Source}

\begin{figure}
\begin{center}\includegraphics[width = 0.52\textwidth]{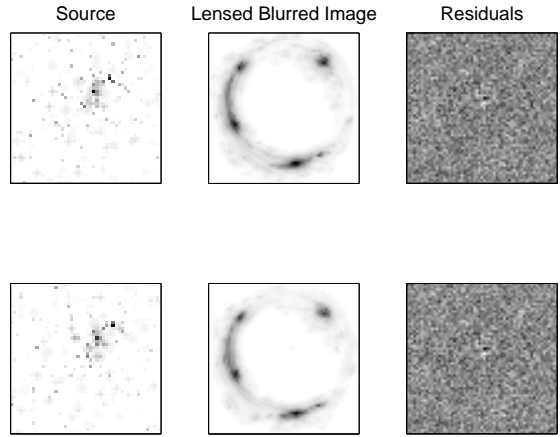}
\end{center}
\caption{Two samples from the posterior distribution of sources, along
with  their lensed  blurred images and  the
normalized residuals between the model  and the data. The source plane
is 0.6 arc seconds across, corresponding to a physical length scale of
$\sim$ 7 kpc.\label{source}}
\end{figure}

Figure~\ref{source}  shows  two   results  from  the  MCMC  simulation
(i.e.  two  samples   from  the  posterior  distribution).  The lensed and blurred images are also shown, and the residuals between the images and the observed image are consistent with pure noise, in terms of $\chi^2$. If the number of degrees of freedom (data points minus effective number of free parameters) was well defined, the image would not match the observed one to the extent that a frequentist goodness of fit test would demand. However, since we are not using optimization, these common tests do not apply.

\begin{figure}
\begin{center}\includegraphics[width = 0.52\textwidth]{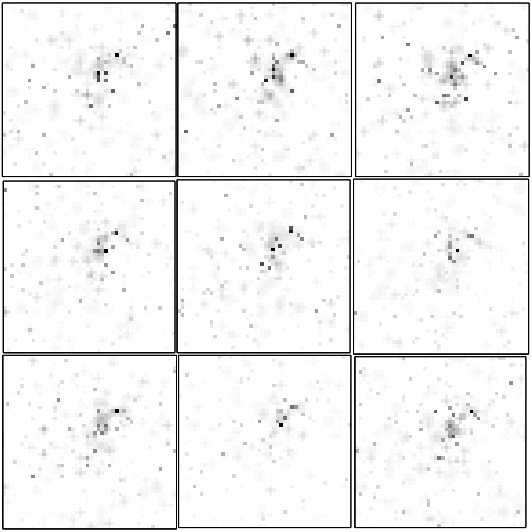}
\end{center}
\caption{More sample sources from the posterior distribution. The diversity across the samples is an indication of the uncertainty that we have about the nature of the source.\label{moresamples}}
\end{figure}

Individual reconstructions  tend to  have their  flux concentrated  into  a small
number of pixels (across a physical length scale of 1-2 kpc\footnote{The assumed cosmological parameters
were  $\Omega_m = 0.27$,  $\Omega_\Lambda =  0.73$ and  H$_0 =  71$ km
s$^{-1}$ Mpc$^{-1}$.}), but the  reliable conclusions are the ones which are
reproduced   across  most   of   the  sources   in   the  sample. Nine more samples are shown in Figure~\ref{moresamples}.

\begin{figure}
\begin{center}\includegraphics[width = 0.5\textwidth]{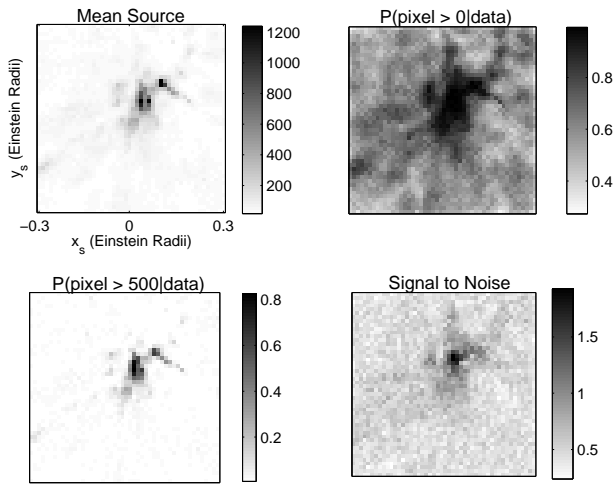}
\end{center}
\caption{The  mean   source  reconstruction (top left) serves as a final estimate of the source, with the two components clearly visible. The next two panels are the marginal posterior probabilities that each pixel is greater than zero, and greater than 500 respectively, and so quantify the amount of evidence for each part of the source. The bottom right plot is the ratio of the mean flux to the standard deviation, showing that we have the most information about the brightest parts of the source.\label{meansource}}
\end{figure}

In Figure~\ref{meansource}, the posterior  mean source is displayed. This
may be taken  as a final estimate of the source  (if a single estimate
is  required),  and when this is lensed, it is consistent with the image to within the size of the noise error bars. The  posterior samples and the mean source both
show strong evidence  for complex structure in the  upper right region
of the  source, with a  component protruding out at right angles  to the
main component. Other authors have suggested that this may be a system
of        interacting       galaxies       \citep{2005ApJ...623...31D,
2005MNRAS.360.1333W}. This feature is present in the vast majority of the 
source samples, indicating that the evidence for two components is strong.

In principle, to decide whether or not this is a double source, we should 
measure the posterior probability that the source is double, by counting 
the fraction of samples which are double. But this is problematic - since 
we are using a nonparametric source model, it becomes hard to define what 
is meant by a double source. Despite this difficulty, the question can be 
answered by simply observing the sequence of source models and noting
that the component at right angles to the main light source is persistent 
across the vast majority of the posterior samples (Figure~\ref{moresamples}), indicating a robust 
conclusion. This is also demonstrated by the significance maps in Figure~\ref{meansource}.

In the primary component of the mean source map, some substructure is visible, suggesting that we may have resolved extra detail in the core of this galaxy. However, the low value of the signal to noise ratio ($\sim$ 1) casts doubt on this conclusion. Individual pixels are highly underconstrained in our approach, it is the integral properties of the source that are robustly measured. As a check, we calculated the posterior probability that both pixels (20,27) and (20,29) are brighter than pixel (20,28), and found this probability to be 0.48, and hence that this question cannot be answered with the current data. Also, there is some weak evidence for more patches of the light to the left of the primary component, but better data would be required to firmly establish this.

\begin{figure*}
\begin{center}\includegraphics[width = 0.8\textwidth]{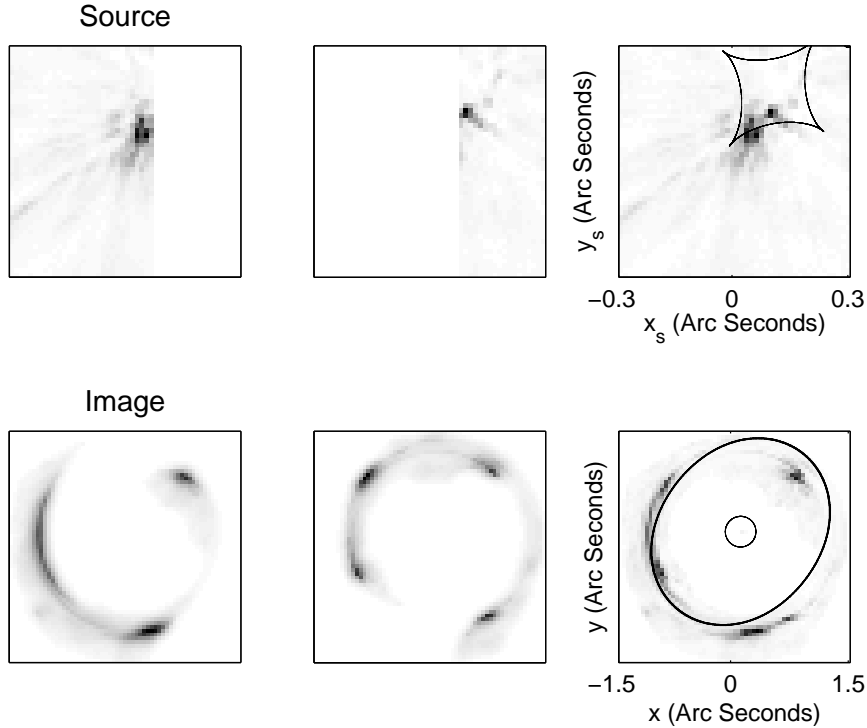}
\end{center}
\caption{The  two  components  of  the reconstructed  source,  plotted
separately to show how they contribute to the image. On the right, the
estimated source and lensed (not blurred) image are plotted along with
the   caustics  and  critical   lines  for   the  best   fitting  lens
model.\label{seahorse}}
\end{figure*}

In  Figure~\ref{seahorse}, the  two  possible components  of the  mean
source are plotted  separately, and the images of  them are shown. The
first component contributes to three of the images, but causes a large
extended  region  of light  on  the left  of  the  image.  The  second
component  is  needed  in   order  to  explain  the  additional,  more
concentrated component of the leftmost  image, as well as the presence
of   the  fourth   image. A   similar  conclusion   was   reached  by
\citet{2005MNRAS.360.1333W} and \citet{2005ApJ...623...31D}, however, our simulations have found evidence for a structure that is long and narrow, on the scale of the pixel size that we have used. Since other authors have used larger pixels, they have found that the best fitting source is a binary source in which both components are blobby, whereas our results suggest a needle-like structure for the second component, with a width of $\sim$ 200 pc.

\subsection{Lens Parameters}
The   estimates  of   the   lens  model   parameters   are  shown   in
Table~\ref{lensparams} (the units, where appropriate, are arcseconds, except for $\theta$ which is in radians).
These     estimates    are     similar    to     those     found    by
\citet{2005MNRAS.360.1333W}  using a  similar (PIEP)  lens  model, but our uncertainties are greater, by a factor of $\sim$ 1-4.

There   are   several   factors   contributing   to   this   increased
uncertainty. One is due to the fact that we have used more pixels than
others,  making  weaker  a   priori  assumptions  about  the  source's
structure. We  have also used  a different, more reasonable  prior for
the source, taking into account the fact that astronomical sources are
usually     localized      bright     patches     with      a     dark
background. Regularization based priors \citep{suyu} may not take this
into account.

Another  possible   cause  for   the  increased  uncertainty   is  the
approximations  that have  previously been  used, such  as  a Gaussian
approximation to the posterior PDF about its peak. This is typically a
good approximation  if the data  constrains the parameters  very well,
but  this is not  usually the  case with  ``non-parametric'' (actually
many-parametric) reconstructions.  Finally, we did not  fix the centre
of the lens mass distribution at a particular point, but included this
as an extra pair of parameters to be inferred from the image, and also included the $\gamma$ and $r_c$ parameters.

To infer any properties of this lensing system (with corresponding 
uncertainties), all that is required is to calculate the desired property 
for all of the posterior samples, and observe the diversity of the 
answers. As an example, we were able to measure 
the total magnification of this system, defined as the flux in the image 
divided by the flux of the source; this value was found to be 17.9 $\pm$ 
1.7, further demonstrating the value of lensing for investigations of high 
redshift galaxies.

The lens mass distribution was found to be slightly elliptical, with q = 0.932 $\pm$
0.006, and  the angle of  orientation of the  lens (which can  be seen
from the orientation of  the critical curves in Figure~\ref{seahorse})
is   close    to   that   of   the    subtracted   foreground   galaxy
\citep[see][]{2005MNRAS.360.1333W}.   

From  these results  we can  also  estimate the  total mass  
contained within  the image.   We defined  a circle  of radius  1.2  arc seconds
centred  in  the  middle  of  the image  and  integrated  the  density
(proportional to  the Laplacian of the lensing  potential) within this
ring, for each lens parameter set. Given our  assumptions, the mass contained within
the ring was  found to be (2.91$\pm$0.01)$\times$10$^{11}$ M$_{\sun}$,
in agreement  with the value  reported by \citet{2003ApJ...583..606K},
who found  a slightly  lower value  but also used  a ring  of slightly
smaller radius.

\begin{table}
\begin{center}
\caption{Results for lens model parameters. Where 
appropriate, the units are arc seconds, except for $\theta$, which is in radians. The values found by \citet{2005MNRAS.360.1333W} for the similar PIEP model are also shown. For the $\gamma$ parameter, the quoted value from \citet{2005MNRAS.360.1333W} was from a softened power-law elliptical mass density model.}\label{lensparams}
\begin{tabular}{lcc}
\hline Parameter & Value & \citet{2005MNRAS.360.1333W}\\
\hline $b$ & 1.177 $\pm$ 0.016 (''$^{1-\gamma}$)& 1.170 $\pm$ 0.004\\
$q$ & 0.932 $\pm$ 0.006 & 0.917 $\pm$ 0.004\\
$r_c$ & 0.09 $\pm$ 0.07 & N/A\\
$\gamma$ & 1.04 $\pm$ 0.06 & 1.08 $\pm$ 0.03\\
$x_c$ & 0.107 $\pm$ 0.006 & N/A\\
$y_c$ & 0.176 $\pm$ 0.008 & N/A\\
$\theta$ & 2.245 $\pm$ 0.009 & 2.247 $\pm$ 0.01\\
$\sigma$ & Negligible & N/A
\end{tabular}
\medskip\\
\end{center}
\end{table}

\section{Conclusions}\label{conclusions}
In this paper,  we have presented an analysis of the  HST image of the
Einstein  ring  0047-2808 using  a  Bayesian  procedure.  Most of  the
results are qualitatively consistent with previous work, leading us to conclude that
the   various   different   methods   available   for   reconstructing
gravitational  lens systems  all  give satisfactory  results for  this
system,  despite  the  disagreement  about the  uncertainties  in  the
parameter estimates.   However, this may  not always be the  case with
other data sets \citep[see][for an example]{2005ApJ...631.1198G}. However, our use of smaller pixels and a more reasonable prior for the source has allowed us to reconstruct the source with a greater resolution than has been possible with other methods. As a result, we have found that the second component of the source is not just a single patch of light, but has a narrow structure protruding from it; given structure formation in $\Lambda$CDM cosmologies, we should expect such objects in the early universe to show significant substructures indicative of major mergers and accretion.

Discovered in continuum radio sources \citep{1988Natur.333..537H}, the
number,  diversity  and observational  detail  of  Einstein rings  has
continued          to         grow~\citep[e.g.][]{2003ApJ...583...67C,
2003MNRAS.343L..29T,    2003Sci...300..773C}.   Furthermore,   optical
Einstein   rings   should  be   uncovered   in   a  upcoming   surveys
\citep{1992MNRAS.259P..31M}, particularly the SLACS survey \citep{2006ApJ...638..703B},  and  the  number  of  cases  is  steadily
increasing \citep{2003A&A...406L..43S,  2005A&A...436L..21C}. Hence it
is vital that progress is made in developing and understanding optimal
and  reliable  methods  for  analyzing  them.   Model  selection  will
probably  play  an   important  role  in  this  area   in  the  future
\citep{suyu}, however we  doubt that it will be  useful to compare the
evidence for  one ad-hoc regularization formula  against another, when
it is  known in advance that  neither of them  accurately describe the
prior information that is  available.  Despite these concerns, the level of qualitative agreement
between the  results presented in this  paper and in  others show that
these debates may only be of minor importance in practice.

\section*{Acknowledgments}
We  would like to  thank Randall  Wayth and  Radford Neal  for valuable
discussion, and  further thank Randall  Wayth for providing  the images
and noise frames used in  this study. We also appreciate the comments of the anonymous referee, which helped to improve the paper. GFL acknowledges support for ARC DP0452239.


\end{document}